\documentclass[conference]{IEEEtran}
\IEEEoverridecommandlockouts

\usepackage{cite}
\usepackage{amsmath,amssymb,amsfonts}
\usepackage{algorithmic}
\usepackage{graphicx}
\usepackage{textcomp}
\usepackage[table]{xcolor}
\def\BibTeX{{\rm B\kern-.05em{\sc i\kern-.025em b}\kern-.08em
    T\kern-.1667em\lower.7ex\hbox{E}\kern-.125emX}}
\definecolor{lightcyan}{rgb}{0.88, 1.0, 1.0}

\usepackage{hyperref}
\definecolor{mattcolor}{HTML}{0004ff}

\usepackage{microtype}

\usepackage{amsmath}
\usepackage{amssymb}
\usepackage{booktabs}
\let\vec\overrightarrow
\DeclareMathOperator*{\argmin}{arg\,min}

\usepackage{moresize}
    
\begin{document}
\makeatletter
\newcommand{\linebreakand}{
  \end{@IEEEauthorhalign}
  \hfill\mbox{}\par
  \mbox{}\hfill\begin{@IEEEauthorhalign}
}
\makeatother

\title{\huge kNN-SVC: Robust Zero-Shot Singing Voice Conversion with Additive Synthesis and Concatenation Smoothness Optimization\\
\thanks{This project is partially supported by the European Research Council under Europe's Horizon 2020 program, grant \#883313 (ERC REACH).}
}

\author{\IEEEauthorblockN{Keren Shao}
\IEEEauthorblockA{
\textit{University of California San Diego, La Jolla, USA}\\
k5shao@ucsd.edu}
\and
\IEEEauthorblockN{Ke Chen}
\IEEEauthorblockA{
\textit{University of California San Diego, La Jolla, USA}\\
knutchen@ucsd.edu}
\linebreakand
\hspace{-0.25cm}
\IEEEauthorblockN{Matthew Baas}
\IEEEauthorblockA{
\hspace{-0.25cm}\textit{Stellenbosch University, Stellenbosch, South Africa}\\
\hspace{-0.25cm}20786379@sun.ac.za}
\and
\hspace{-0.25cm}
\IEEEauthorblockN{Shlomo Dubnov}
\IEEEauthorblockA{
\hspace{-0.25cm}
\textit{University of California San Diego, La Jolla, USA}\\
\hspace{-0.25cm}
sdubnov@ucsd.edu}
}

\maketitle

\begin{abstract}

    Robustness is critical in zero-shot singing voice conversion (SVC). This paper introduces two novel methods to strengthen the robustness of the kNN-VC framework for SVC. First, kNN-VC's core representation, WavLM, lacks harmonic emphasis, resulting in dull sounds and ringing artifacts. To address this, we leverage the bijection between WavLM, pitch contours, and spectrograms to perform additive synthesis, integrating the resulting waveform into the model to mitigate these issues. Second, kNN-VC overlooks concatenative smoothness, a key perceptual factor in SVC. To enhance smoothness, we propose a new distance metric that filters out unsuitable kNN candidates and optimize the summing weights of the candidates during inference. Although our techniques are built on the kNN-VC framework for implementation convenience, they are broadly applicable to general concatenative neural synthesis models. Experimental results validate the effectiveness of these modifications in achieving robust SVC. Demo: \href{http://knnsvc.com}{http://knnsvc.com}. Code: \href{https://github.com/SmoothKen/knn-svc}{https://github.com/SmoothKen/knn-svc}
\end{abstract}


\section{Introduction}
\label{sec:intro}

Singing voice conversion (SVC)—the process of transforming a source singer’s voice to match a reference voice with pitch adjustments while preserving lyrical content—has recently benefited from advancements in neural networks~\cite{vc-survey-2021, huang2023singing, hussain2023ace, unsupervised-svc, pitchnet, fastsvc}. Some prior approaches, such as CycleGAN~\cite{cycleganvc} and StarGAN~\cite{StarGANv2VC}, achieve implicit disentanglement via cycle consistency loss, while others like diffsvc~\cite{liu2021diffsvc}, ppg-svc~\cite{ppg-svc}, \textit{so-vits-svc}\footnote{\href{https://github.com/svc-develop-team/so-vits-svc}{https://github.com/svc-develop-team/so-vits-svc}} and \textit{DDSP-SVC}\footnote{\href{https://github.com/yxlllc/DDSP-SVC}{https://github.com/yxlllc/DDSP-SVC}} attempt explicit disentanglement using PPG and pretrained ContentVec~\cite{contentvec} features. However, even with carefully designed models, fully separating content from style remains a challenge~\cite{soft-vc-2022}, often resulting in unclear articulation or unintended timbre leakage in converted audio, which can compromise the robustness of SVC systems.

\begin{figure}[t]
  \centering
  \includegraphics[width=\linewidth]{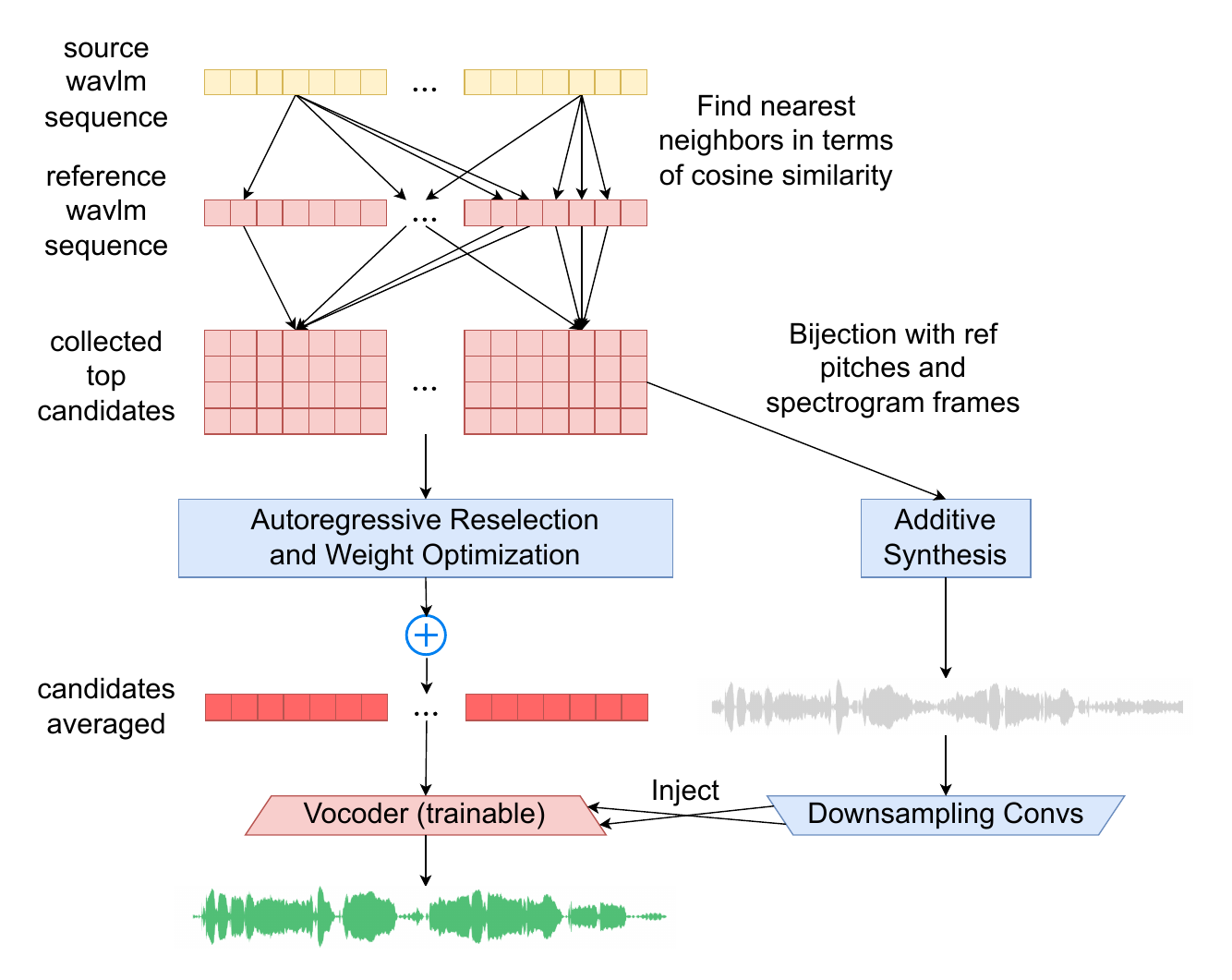}
  \caption{Workflow of the kNN-SVC model: The left blue block represents Concatenation Smoothness Optimization, while the Additive Synthesis pipeline occupies the blue column on the right. The original kNN-VC backbone is composed of the remaining red blocks.}
  \label{fig:knnvc}

\end{figure}

Concatenative neural synthesis models, such as kNN-VC~\cite{baas2023voice} and \textit{RVC}\footnote{\href{https://github.com/RVC-Project/Retrieval-based-Voice-Conversion-WebUI}{https://github.com/RVC-Project/Retrieval-based-Voice-Conversion-WebUI}}, bypass the disentanglement challenge by using non-parametric nearest neighbor regression within a self-supervised learning (SSL) representation space~\cite{WavLM, HuBERT}. This approach yields audio with high timbre similarity to the reference speaker while maintaining strong intelligibility. To extend this robustness to SVC, we derive two key insights from kNN-VC and its underlying SSL representation, WavLM.

First, compared to speech, our perception of timbre in singing places much greater emphasis on the $f_0$ and its harmonics~\cite{guoharmonics, kshaoharmonics}. However, WavLM's representation does not adequately capture these factors, due to the nature of its training dataset and pipeline. To address this, we introduce additional harmonic information via additive synthesis, leveraging the bijective relationship between WavLM frames and the corresponding spectrogram frames of the audio.

The second insight concerns the well-known issue of smoothness in concatenative synthesis. In singing, notes can sustain for several seconds, often representing climactic moments in a piece. However, the frame-by-frame nature of kNN-VC leads to candidates optimized for local fidelity without considering their temporal coherence, resulting in artifacts such as slurring or trembling. To mitigate this, we design a novel distance function that incorporates the cost of temporal concatenation, and we use it to autoregressively replace temporally unfit candidates. When combining candidates, we further optimize their weights to minimize the concatenation cost, leading to smoother and more perceptually pleasing output.

In this paper, we propose two techniques to address the challenges of insufficient harmonic representation and temporal smoothness in neural concatenative synthesis models. Our contributions are as follows:

\begin{enumerate}
    \item We introduce harmonic information into the kNN-VC model via additive synthesis, ensuring the authenticity of the speaker’s timbre in a non-parametric manner.
    
    \item We propose a novel distance metric that extends the kNN-VC’s cosine similarity by incorporating a temporal concatenation cost. By autoregressively replacing kNN candidates and optimizing their combination weights during inference, we achieve smoother and more perceptually coherent results.
    
    \item Through both objective evaluation and subjective assessment, we demonstrate that our techniques not only enhance the robustness of singing voice conversion but also improve performance on speech conversion tasks when compared to kNN-VC.
\end{enumerate}

\section{kNN-SVC}

\subsection{kNN-VC overview}

As shown in Figure \ref{fig:knnvc}, we first convert both the source and reference audio into their WavLM representations. At each time step $t$, we find the nearest neighbors of the source WavLM element from the reference pool. We retain the top $k$ neighbors (with $k = 4$, as in the original kNN-VC paper), calculate their mean, and insert this averaged vector at time step $t$ in the concatenated WavLM sequence. The vocoder, which is the only parametric model in this framework, is then trained to generate the final waveform from this concatenated sequence. Since HiFi-GAN requires supervised learning, we use other utterances from the same speaker as references during training.


\subsection{Additive Synthesis}

To inject harmonic information via additive synthesis (AS), we require the pitch and harmonic amplitudes at each time step, \( t \). This can be achieved by noting that each WavLM candidate corresponds to a pitch and a spectrogram frame, assuming the hop size for the pitch extractor and STFT is set to match that of WavLM. Given both the pitch and spectrum, the harmonic amplitudes can be derived by identifying the relevant frequency indices, as shown in Figure \ref{fig:AS}. Consequently, each WavLM candidate is associated with a pitch (or zero if unpitched) and a corresponding vector of harmonic amplitudes.

In SVC, the desired pitch contour is obtained by shifting the source utterance's pitches to fit the reference speaker’s vocal range. Instead of selecting candidates from the concatenated WavLM sequence, we must identify those whose pitch closely matches the desired pitch for additive synthesis. As illustrated in Figure \ref{fig:AS}, we begin with the nearest neighbor candidates, but with a higher cutoff, $k' > k$. From this refined pool, we rank the candidates based on their pitch proximity to the desired pitch, then select the top-ranked candidates.

Using these selected candidates, we first upsample both the desired pitch sequence and the averaged harmonic amplitudes (across candidates) to match the time resolution of the waveform, similar to the approach in the DDSP paper ~\cite{DDSP}. Let these be denoted as \( f_0(t) \) and \( \{A_n\}_{n=1}^N(t) \), respectively. The additive synthesis waveform, \( U(t) \), is then generated as:

\[
U(t) = \sum_{n=1}^N A_n(t) \sin(\text{cumsum}(2\pi n f_0(t)))
\]

Finally, we integrate this synthesized waveform into the original HiFi-GAN model by employing downsampling convolutional layers, as shown in Figure \ref{fig:knnvc}.

\begin{figure}[t]
  \centering
  \includegraphics[width=\linewidth]{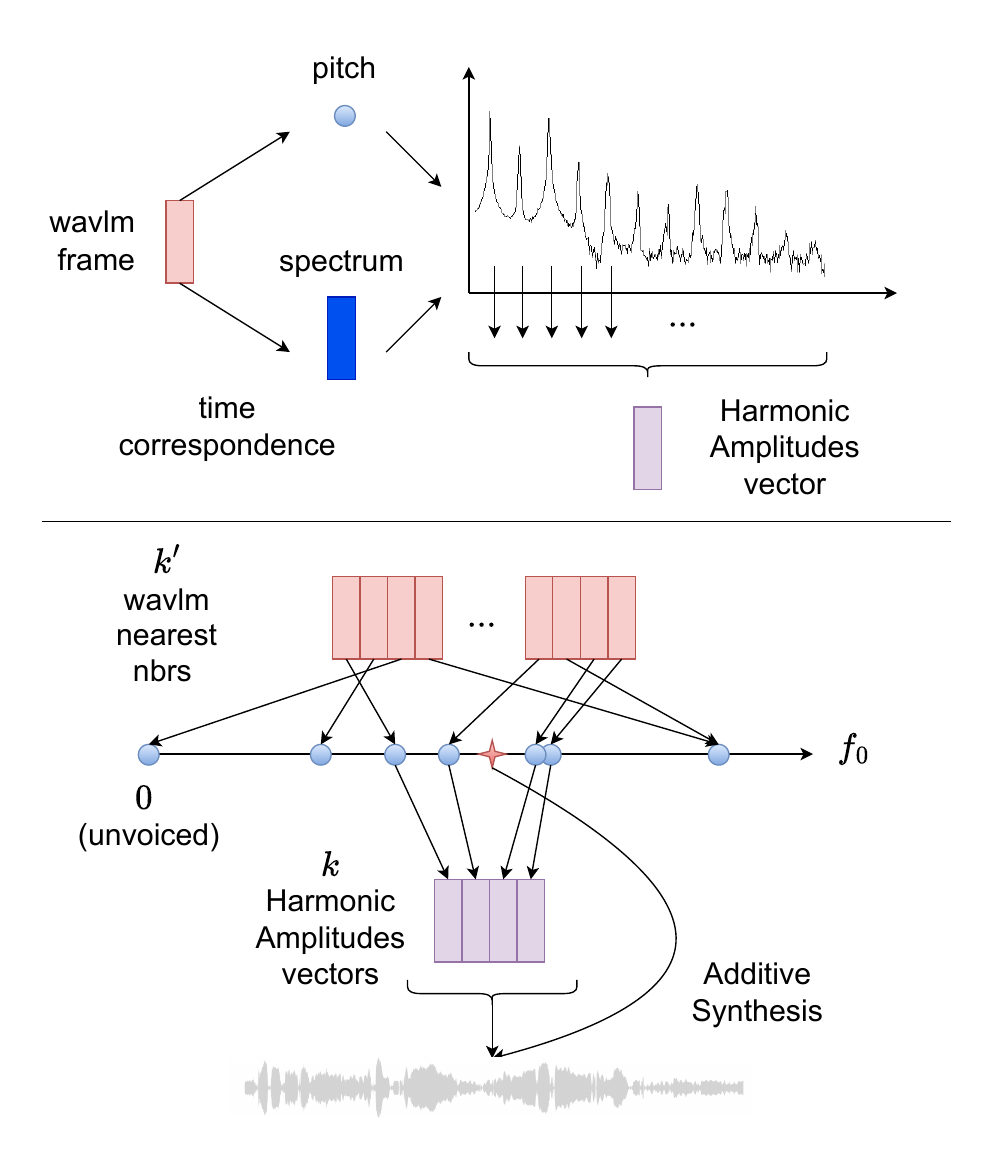}
  \caption{The process of creating additively synthesized waveform. We first extract the corresponding harmonic amplitude vector for each reference WavLM frame (top). With the orange star indicating the target pitch, we then select candidate frames with the closest pitches to perform additive synthesis (bottom).}
  \label{fig:AS}
\end{figure}

\subsection{Concatenation Smoothness Optimization}

\begin{figure}[t]
  \centering
  \includegraphics[width=\linewidth]{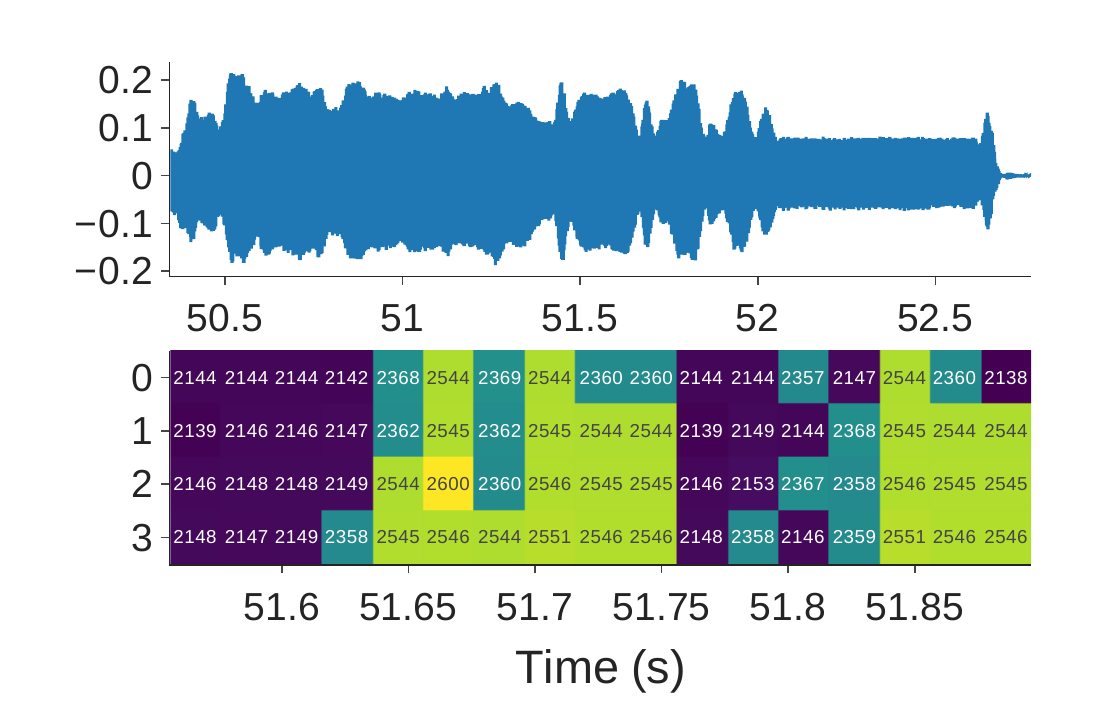}

  \caption{The problem that inference-time Concatenative Smoothness Optimization attempts to address. The numbers in the matrix represent candidate indices from the reference utterance. The presence of short, mutually exclusive territories in the matrix is the primary cause of the trembling artifacts seen in the waveform above.}

  \label{fig:rough_demo}
  
\end{figure}

As illustrated in Figure \ref{fig:rough_demo}, the roughness in the output audio often stems from a sole focus on local cosine similarity. In this instance, with a kNN cutoff at $k = 4$, imperceptible amplitude tides in the source audio lead to short candidate segments whose cosine similarities are temporally inconsistent with their surrounding context. Increasing the cutoff to address this issue would be counterproductive, as it would introduce candidates of lower quality indiscriminately.

To tackle this problem, we propose an autoregressive candidate reselection method, as shown in Figure \ref{fig:CAT}, using the following distance measure:

\
\begin{align*}
\mathcal{L}_{total}(C, t) &= \mathcal{L}_{src}(C, t) + m\mathcal{L}_{concat}(C, t) \\
&= \text{cosine\_sim}(C, S_t) \\
&+ m\text{median}(\{\text{cosine\_sim}(C, C') \text{ for } C' \in \mathcal{A}_{t-1}\})
\end{align*}
\

Here, $t$ represents the current time step, $C$ is an arbitrary WavLM candidate, $S_t$ denotes the source WavLM, and $\mathcal{A}_{t-1}$ is the set of reselected candidates from the previous step. At $t = 0$, we simply use the kNN candidates. To reduce computational complexity for $\mathcal{L}_{concat}$, the new candidates are restricted to frames that serve as continuations (in the reference utterance) of the reselected candidates from the prior step. The hyperparameter $m \geq 0$ balances local accuracy with smoothness. Setting $m = 0$ reduces the method to the original kNN approach.

To further improve robustness in concatenation, we replace the simple averaging of WavLM candidates with a weighted approach. The key insight is that the output can match the smoothness of the reference utterance if the concatenated neighbors of a candidate are exactly its temporal continuation in the reference. As depicted in Figure \ref{fig:CAT}, for candidate sets $\mathcal{A}_{t_0}$ and $\mathcal{A}_{t_1}$, where $t_1 = t_0 + 1$, the final weighted vector at $t_0$ is given by $V_{t_0} = \vec{w_{t_0}} \cdot [C_i \mid i \in \mathcal{A}_{t_0}]$. The ideal right continuation is $R_{t_0} = \vec{w_{t_0}} \cdot [C_{i+1} \mid i \in \mathcal{A}_{t_0}]$, where $i$ is the index of $C_i$ in the reference utterance.

Likewise, at $t_1$, we define $V_{t_1} = \vec{w_{t_1}} \cdot [C_j \mid j \in \mathcal{A}_{t_1}]$, with the ideal left continuation given by $L_{t_1} = \vec{w_{t_1}} \cdot [C_{j-1} \mid j \in \mathcal{A}_{t_1}]$. To optimize, we aim for $V_{t_1}$ to match $R_{t_0}$ from the perspective of $t_0$ and for $V_{t_0}$ to match $L_{t_1}$ from the perspective of $t_1$. Generalizing this for all $t$ and using MSELoss, we minimize the following:

\
\begin{equation*}
\begin{aligned}
& \argmin_{w}  \quad \sum_{t \geq 1} || L_{t} - V_{t - 1} ||_2^2 + || R_{t - 1} - V_{t} ||_2^2 \\
& \text{subject to} \quad || \vec{w_{t}} ||_1 = 1, \quad \vec{w_{t}} \geq 0 \quad \forall t \\
\end{aligned}
\end{equation*}
\

\begin{figure}[t]
  \centering
  \includegraphics[width=\linewidth]{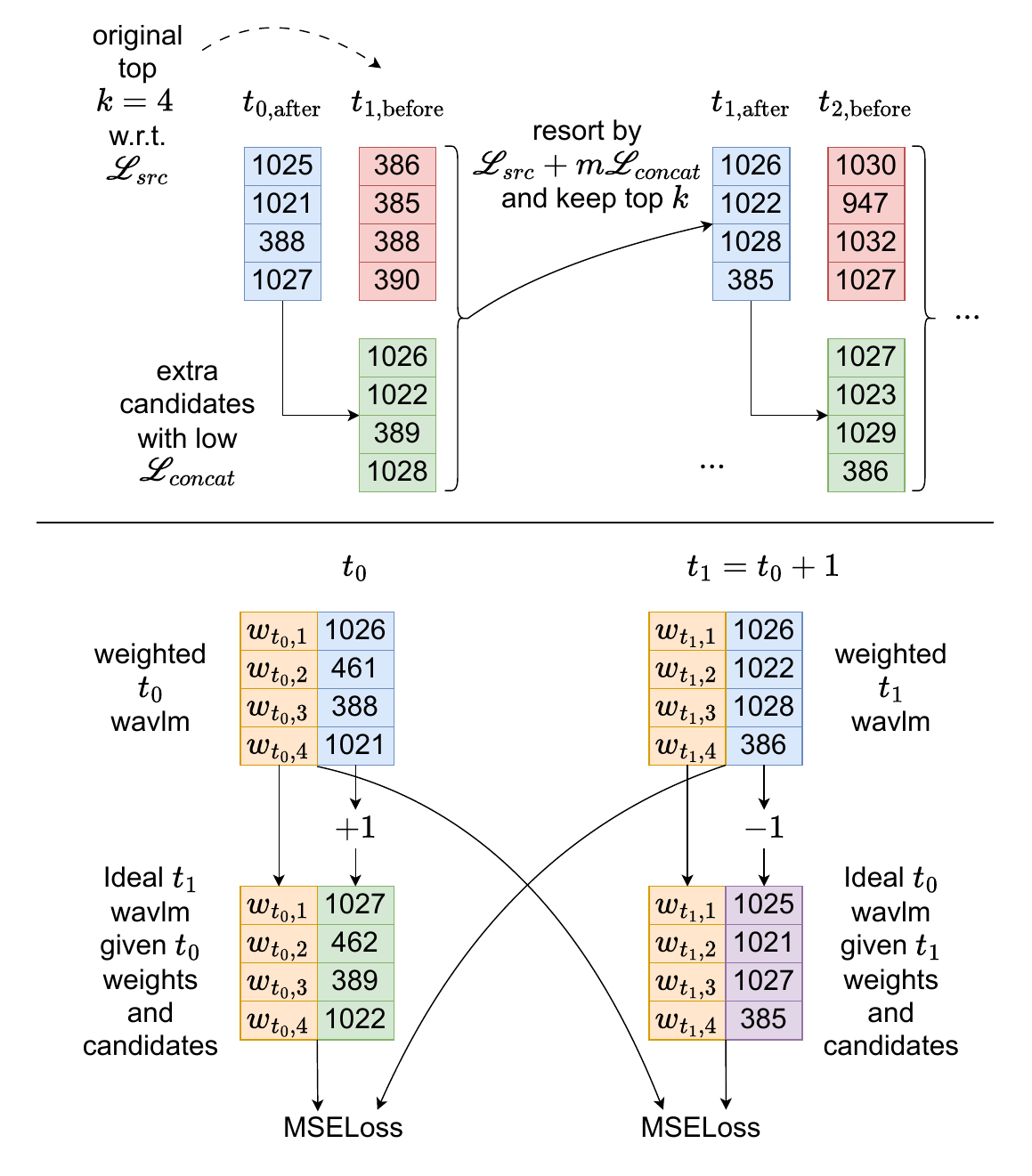}

  \caption{The process of inference-time Concatenation Smoothness Optimization. The numbers represent candidate indices from the reference utterance. Top: Autoregressively reselecting candidates based on a weighted sum of cosine similarity $L_{src}$ and concatenative cost $L_{concat}$. Bottom: Optimizing the summing weights towards minimizing the discrepancy between one's concatenation neighbors and its ideal continuations in the reference utterance.}
  \label{fig:CAT}

\end{figure}

\section{Experiments}
\begin{table*}[t]
\caption{Ablation studies and model comparisons were conducted on the LibriSpeech (LS), OpenSinger (OS), and NUS48E datasets. "AS" refers to the use of Additive Synthesis, while "CAT" denotes inference-time Concatenation Smoothness Optimization. Two sets of kNN-SVC checkpoints were employed: the first, trained on the LibriSpeech training subset, was used exclusively for LS $\to$ LS evaluations, while the second, trained on a subset of OpenSinger, was dedicated to OS $\to$ OS and OS $\to$ NUS48E evaluations. Each objective evaluation presented in the table reflects the mean performance across more than 7,000 converted utterances. Given this large data size, the confidence intervals are approximately 0.05\%, and are therefore omitted from the table. so-vits-svc and DDSP-SVC are incompatible with zero-shot settings due to their use of speaker IDs. A \colorbox{lightcyan}{partial comparison} was performed by including 30\% of the test speaker data in their training.}

\vspace{0.3cm}

\renewcommand\arraystretch{1.3} 

\resizebox{\textwidth}{!}{
\setlength{\tabcolsep}{0.85mm}{
\centering
\begin{tabular}{l|ccccc|ccc|ccc}
\toprule
\multicolumn{1}{c|}{Dataset} & \multicolumn{5}{c|}{LS $\to$ LS} & \multicolumn{3}{c|}{OS $\to$ OS} & \multicolumn{3}{c}{OS $\to$ NUS48E} \\
\cmidrule{1-2} \cmidrule{3-12}
\multicolumn{1}{c|}{Metrics}
& WER$\downarrow$ & CER$\downarrow$ & EER$\uparrow$ & MOS & SIM & EER$\uparrow$ & MOS & SIM & EER$\uparrow$ & MOS & SIM  \\
\cmidrule{1-2} \cmidrule{3-12} 
kNN-VC \cite{baas2023voice} & 5.42 & 2.02 & 38.97 & 3.98 $\pm$ 0.10 & 2.86 $\pm$ 0.13 & - & - & - & - & - & - \\

kNN-SVC w/o \{AS, CAT\} & \textbf{5.27} & \textbf{1.96} & 36.40 & 3.96 $\pm$ 0.11 & 2.89 $\pm$ 0.09 & 48.58 & 3.56 $\pm$ 0.11 & 1.98 $\pm$ 0.10 & 23.68 & 3.45 $\pm$ 0.14 & 2.05 $\pm$ 0.10 \\

NeuCoSVC \cite{Sha2023NeuralCS} & - & - & - & - & - & 49.77 & 3.70 $\pm$ 0.11 & 2.07 $\pm$ 0.08 & 22.53 & 3.49 $\pm$ 0.10 & 1.98 $\pm$ 0.09 \\

\rowcolor{lightcyan}
so-vits-svc & - & - & - & - & - & 39.36 & 3.66 $\pm$ 0.15  & 	1.76 $\pm$ 0.14 & 14.11 &  3.41 $\pm$ 0.16  &  1.52 $\pm$ 0.13 \\

\rowcolor{lightcyan}
DDSP-SVC & - & - & - & - & - & 34.25 & 3.31 $\pm$ 0.15	& 1.71 $\pm$ 0.14 & 9.40 &  3.04 $\pm$ 0.17 & 	1.26 $\pm$ 0.14 \\

kNN-SVC w/o \{CAT\} & 5.40 & 2.11 & 40.97 & 4.10 $\pm$ 0.09 & \textbf{3.04 $\pm$ 0.08} & 50.55 & 3.92 $\pm$ 0.08 & 2.36 $\pm$ 0.09 & 28.37 & 3.62 $\pm$ 0.09 & 2.42 $\pm$ 0.11 \\

kNN-SVC & 6.02 & 2.40 & \textbf{43.79} & \textbf{4.16 $\pm$ 0.10} & 3.02 $\pm$ 0.08 & \textbf{51.39} & \textbf{4.10 $\pm$ 0.13} & \textbf{2.59 $\pm$ 0.09} & \textbf{29.94} & \textbf{3.81 $\pm$ 0.12} & \textbf{2.57 $\pm$ 0.10} \\
\midrule

Testset topline & 4.36 & 1.52 & - & 4.21 $\pm$ 0.08 & 3.26 $\pm$ 0.11 & - & 4.39 $\pm$ 0.10 & 3.04 $\pm$ 0.11 & - & 4.43 $\pm$ 0.11 & 3.15 $\pm$ 0.07 \\

\bottomrule

\end{tabular}}}

\label{tab:aba_test}
\end{table*}

\subsection{Datasets and Experiment Setup}

\begin{figure}[t]
  \centering
  \includegraphics[width=0.95\linewidth]{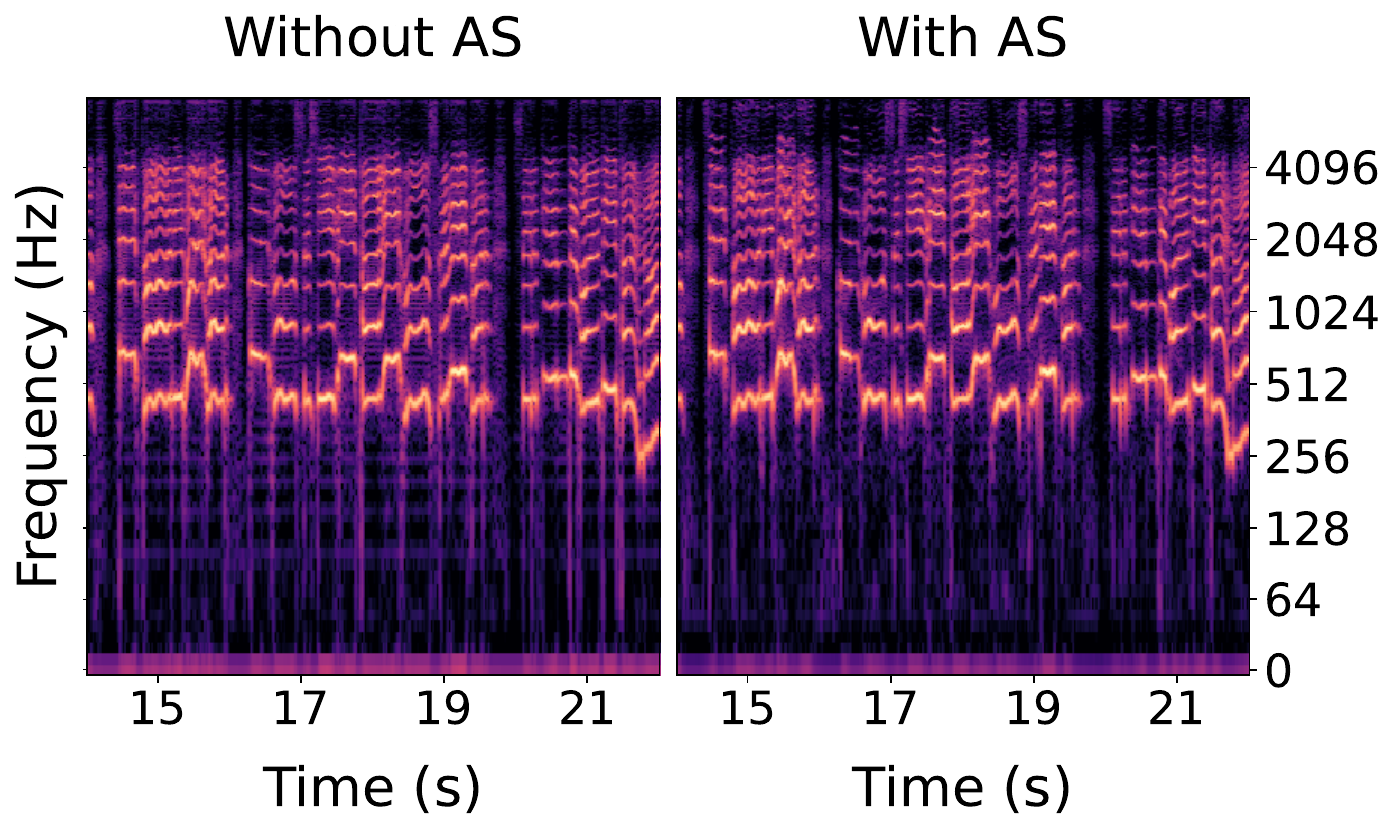}
  \caption{Left: Spectrogram of the output without Additive Synthesis, showing a dull voice quality (due to the absence of high-frequency harmonics) and noticeable ringing artifacts (horizontal dark-blue lines between harmonics). Right: Spectrogram of the output with Additive Synthesis, demonstrating improved harmonic richness and a cleaner signal.}
  \label{fig:AS_visual}
\end{figure}

\begin{figure}[t]
  \centering
  \includegraphics[width=0.95\linewidth]{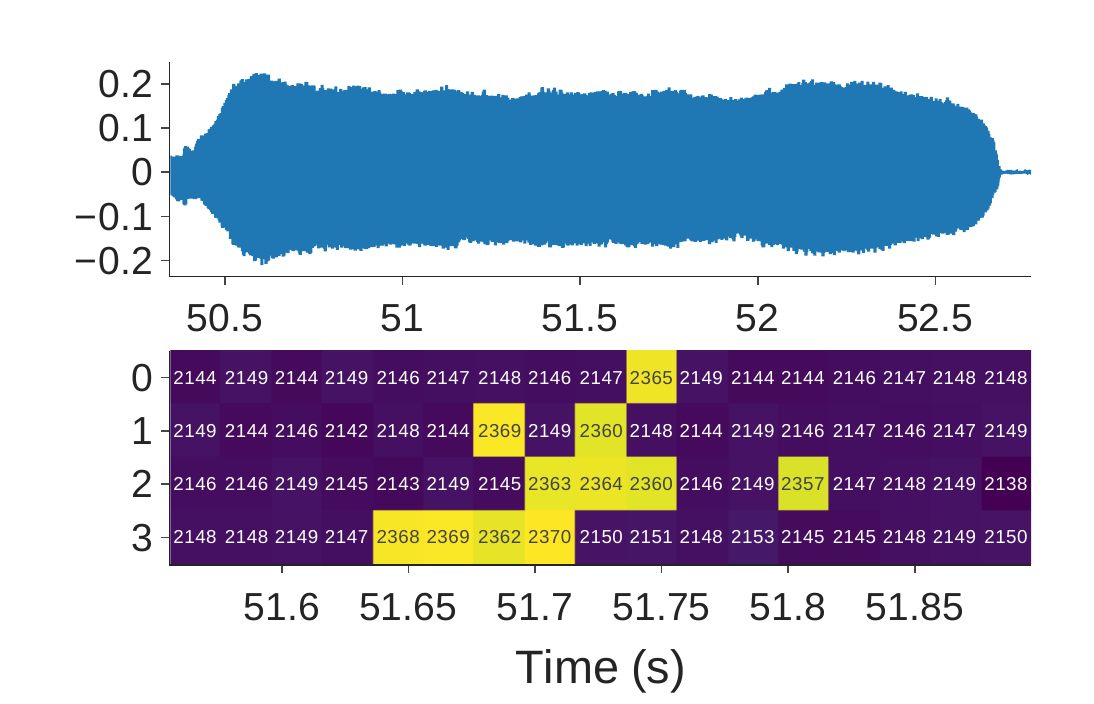}
  \caption{Compared to Figure \ref{fig:rough_demo}, most temporally misaligned WavLM candidates are eliminated after applying autoregressive reselection. While a few may remain, their summing weights approach zero following weight optimization. As a result, the final waveform is significantly smoother.}
  \label{fig:CAT_visual}
\end{figure}

To demonstrate the robustness of our techniques, we compare kNN-SVC with kNN-VC~\cite{baas2023voice} for the speech conversion task, and with NeuCoSVC~\cite{Sha2023NeuralCS} for the singing voice conversion task. NeuCoSVC, an extension of kNN-VC for singing voice conversion, constructs a jointly trained neural harmonic signal generator to produce the input waveform. However, as our experimental results will show, compared to non-parametric additive synthesis, this method lacks robustness in zero-shot SVC.

For speech conversion, we train each of our ablation models on the training subset of the LibriSpeech dataset~\cite{panayotov2015librispeech} and evaluate zero-shot performance by converting among speakers in the test subset.

For singing voice conversion, we reserve two male singers (M26 and M27) and two female singers (F46 and F47) from the OpenSinger dataset~\cite{opensinger} for intra-language evaluation. The remaining data is used for training. For cross-language evaluation, we convert the reserved singers to JEEE, MPUR, MCUR, and SAMF from the NUS48E dataset~\cite{nus48e}. We set the hyperparameters as follows: $k' = 32$, $N = 50$, and $m = 0.3$.

We adopt the WER, CER, and EER metrics from kNN-VC for objective evaluation, and the MOS (scale 1-5) and SIM (scale 1-4) scores for subjective evaluation. In each task (LS $\to$ LS, OS $\to$ OS, and OS $\to$ NUS48E), we sample 60 converted utterances (20 source segments, each converted to 3 different target speakers, using all reference speaker utterances for pre-matching). Study participants were asked to provide MOS and SIM ratings for each conversion pair. Eighteen participants, fluent in both English and Chinese, took part in the subjective evaluations.

\subsection{Model Comparisons and Improvement Visualization}

Our results, presented in Table \ref{tab:aba_test}, clearly demonstrate the advantages of our techniques in terms of EER (objective speaker similarity), as well as MOS and SIM scores. Notably, the trade-offs in WER and CER (word/character error rates) are minimal and largely imperceptible to human listeners.

It is also worth highlighting that, while the parametric synthesizer used by NeuCoSVC begins to falter when transitioning from the OS $\to$ OS task to the more challenging OS $\to$ NUS48E case, our non-parametric additive synthesis, combined with inference-time optimization, remains robust across both objective and subjective metrics.

Finally, the effectiveness of Additive Synthesis and Concatenation Smoothness Optimization is visually illustrated in Figures \ref{fig:AS_visual} and \ref{fig:CAT_visual}, respectively.

\bibliographystyle{IEEEbib}
\bibliography{strings,refs}

\end{document}